\documentclass[twocolumn,showpacs,preprintnumbers,amsmath,amssymb]{revtex4}

\usepackage{graphicx}
\usepackage{dcolumn}
\usepackage{bm}
\usepackage{epsfig}

\begin{document}

\title{Fermionization and fractional statistics\\ in the
strongly interacting one-dimensional Bose gas} 
\author{M.T. Batchelor and
  X.-W. Guan} \affiliation{Department of Theoretical Physics, Research
  School of Physical Sciences and Engineering, and\\ Mathematical
  Sciences Institute, Australian National University, Canberra ACT
  0200, Australia}

\date{\today}

\begin{abstract}
\noindent
We discuss recent results on the relation between the strongly interacting one-dimensional 
Bose gas and a gas of ideal particles obeying nonmutual generalized exclusion statistics (GES).
The thermodynamic properties considered include the statistical profiles, the specific heat
and local pair correlations. 
In the strong coupling limit $\gamma \to \infty$, the Tonks-Girardeau gas, the equivalence
is with Fermi statistics.
The deviation from Fermi statistics during boson fermionization for finite but large 
interaction strength $\gamma$ is described by
the relation $\alpha \approx 1 - 2/\gamma$, where $\alpha$ is a measure of the GES.
This gives a quantitative description of the fermionization process.
In this sense the recent experimental measurement of local pair correlations in a
1D Bose gas of $^{87}$Rb atoms also provides a measure of the deviation of the GES parameter $\alpha$ 
away from the pure Fermi statistics value $\alpha=1$.
Other thermodynamic properties, such as the distribution profiles and the specific heat, 
are also sensitive to the statistics. 
They also thus provide a way of exploring fractional statistics in the 
strongly interacting 1D Bose gas. 
\end{abstract}



\maketitle

The one-dimensional (1D) Bose gas with delta-function interaction \cite{LL,McGuire} 
is an elegantly simple exactly solved quantum many-body system  
which is testable in experiments on trapped quantum gases of ultracold 
atoms \cite{E-TG1,E-TG2,E-TG3,E-TG4,E-TG5,Weiss1,Weiss2}.  
The model has been extensively studied 
(see reviews  \cite{MAT,Gaudin,KOR,Tbook,CAZA1,LUTT1,Yukalov1,Yukalov2}). 
The novel quantum many-body effects inherent in the 1D interacting Bose gas continue 
to attract theoretical and experimental interest 
\cite{Weiss1,Weiss2,Che,Damski,Caux,Schmidt,Forrester,Campo,Buljian,Yu,Rigol}.

In the experiments an effectively 1D system of bosons is made by 
tightly confining the atomic gas in two radial directions and
weakly confining it along the axial direction.
In this way the motion of particles along the radial directions 
is frozen out, thus making the system effectively 1D. 
The delta-function interaction between bosonic atoms
can be realized via short range interaction with an effective coupling
constant $g_{\rm 1D}$ determined through an effective 1D scattering
length $a_{\rm 1D}$ \cite{Olshanii}. 
In terms of the atomic mass $m$ the dimensionless coupling constant 
\begin{equation}
\gamma =\frac{mg_{\rm 1D}}{\hbar^2 n}, 
\end{equation}
is varied to characterize different interaction regimes by tuning the coupling strength
$g_{\rm 1D}$ or the particle number density $n$. 
In the weak coupling regime $\gamma \ll 1$, the 1D Bose gas
can undergo a quasi-Bose-Einstein condensation due to the coherence of
individual particle wave functions. 
In the strong coupling regime $\gamma \gg 1$, the bosons behave like 
impenetrable hard core particles, known as the Tonks-Girardeau gas (TG) \cite{TG}.

The experimental realization of the 1D TG gas
\cite{E-TG1,E-TG2,E-TG3,E-TG4,E-TG5,Weiss1} has provided significant
insight into the  ``fermionization" of interacting bosons. 
During the fermionization process the local pair correlations of strongly
interacting bosons gradually vanish as the interaction strength $g_{\rm 1D}$
increases \cite{Shlyapnikov,Forrester2,CAZA2,Olshanii2,Astrak}.
This behaviour was observed clearly in the local pair correlation function obtained 
from measurements of photoassociation rates in a 1D Bose gas of $^{87}$Rb atoms  \cite{Weiss1}.
In particular, the wave functions of two colliding particles tolerate 
a degree of overlap for finitely strong interaction.  
This suggests that the behaviour of strongly interacting bosons
deviates from pure Fermi behaviour for finite interaction strength. 
In this sense the quasiparticle excitations of strongly
interacting bosons confined in 1D do not obey pure Fermi statistics.

In 1D the dynamical interaction and the
statistical interaction are inextricably related due to the pairwise
interaction between particles. 
As a result 1D interacting systems can map to a system of ideal particles 
obeying Haldane exclusion statistics \cite{Haldane,Wu}. 
Haldane formulated a general description of quantum
statistics based on generalized Pauli exclusion statistics,
now called generalized exclusion statistics (GES).  
For GES the dimensionality of the single particle Hilbert space depends linearly 
on the particle numbers of other species when certain species of particles are added. 
Soon after the formulation of GES, Wu \cite{Wu}  found that the most probable distribution is
determined by GES through counting the dimensionality of the single particle Hilbert space. 
This definition allows different species to refer to identical particles with different quasimomenta in a wide
range of 1D integrable models \cite{Wu}. 
Over a decade ago it was shown that the 1D Calogero-Sutherland model \cite{Ha} and 
the 1D interacting Bose gas \cite{Wu,Isakov} are equivalent to an ideal gas obeying GES.
This equivalence has been recently investigated for a 1D model of interacting anyons \cite{BGO}.
An anyon-fermion mapping has been proposed to obtain 
solutions for several models of ultracold gases with 1D anyonic exchange symmetry \cite{Girardeau}.

Here we discuss the statistical profiles and the
thermodynamic properties of the strongly interacting 1D Bose gas through
the GES and thermodynamic Bethe ansatz (TBA) \cite{Yang-Yang} approaches. 
The results presented here are a special case of results \cite{BGO,BG} obtained 
recently for the more general 1D model of interacting anyons \cite{Kundu,BGO}, 
which contains the 1D interacting Bose gas as a special case.
Here we highlight and discuss these results in the context of the strongly interacting 1D Bose gas.
At low temperatures the quantum statistics of strongly interacting bosons confined
in 1D is nonmutual, i.e., the GES parameter introduced in the next Section is given by
$\alpha_{i,j}=\alpha\,\delta _{i,j}$, which also defines $g$-on or 
fractional statistics \cite{Wilczek,Sen,Khare}. 
The strongly interacting 1D Bose gas implements a continuous range of GES,  
approaching Fermi statistics with $\alpha=1$ as the interaction strength $g_{\rm 1D} \to \infty$.  
We discuss the total energy, specific heat and local pair correlations in terms of the interaction
strength and the temperature. 
The results indicate that experiments on the strongly interacting 1D Bose gas 
may be used as a testing ground for observing fractional statistics.

\section{Bethe Ansatz and GES}
\label{model}

The 1D model of $N$ interacting bosons is defined by the Hamiltonian \cite{LL,McGuire}
\begin{equation}
H_N=-\frac{\hbar ^2}{2m}\sum_{i = 1}^{N}\frac{\partial
^2}{\partial x_i^2}+\,g_{\rm 1D}\sum_{1\leq i<j\leq N} \delta (x_i-x_j),
\label{Ham1}
\end{equation}
under periodic boundary conditions. 
The coupling constant is determined by $g_{\rm
1D}={\hbar^2c}/{m}={2\hbar ^2}/{(m|a_{\rm 1D}|)}$ where the coupling
strength $c$ is tuned through an effective $1$D scattering length
$a_{\rm 1D}$ via confinement \cite{Olshanii}.
Hereafter we set $\hbar =2m=1$ for convenience. 
We shall restore physical units in the thermodynamics later.
The energy eigenvalues are given by $E=\sum_{j=1}^N k_j^2$, where the
individual quasimomenta $k_j$ satisfy the Bethe equations \cite{LL}
\begin{equation}
{\mathrm e}^{\mathrm{i}k_jL}=- \prod^N_{\ell = 1} 
\frac{k_j-k_\ell+\mathrm{i}\, c}{k_j-k_\ell-\mathrm{i}\, c},  
\label{BA1}
\end{equation}
for $j = 1,\ldots, N$.  
Here $L$ is the length of the system with $n=N/L$ the particle density.

In the thermodynamic limit, $L \to \infty$ with $n$ finite, 
the ground state energy per particle is given by
$E/N=n^2e(\gamma)$ where 
\begin{equation}
e(\gamma)=\frac{\gamma^3}{\lambda^3}\int_{-1}^1\rho(x)x^2dx.
\end{equation}
The quasimomentum distribution function $\rho(x)$ and the parameter
$\lambda =c/Q$, where $Q$ is the cut-off momentum at zero temperature, 
are determined by the integral equation \cite{LL}
\begin{eqnarray}
 \rho(x)&=&\frac{1}{2\pi}+\frac{\lambda}{\pi}\int_{-1}^{1}
 \frac{\rho(y)dy}{\lambda^2+(x-y)^2}, \nonumber\\
  \lambda&=& \gamma \int_{-1}^1\rho(x)dx. 
\label{BA2}
\end{eqnarray}
In the strong and weak coupling limits, the leading terms in the 
ground state energy can be obtained analytically from the Bethe equations (\ref{BA1})
\cite{Gaudin,BGM,Forrester} or (\ref{BA2}) \cite{LL,Wadati}.
In the strong coupling regime $\gamma \gg 1$ the ground state energy per particle is given by
\begin{equation}
\frac{E_0}{N} \approx
\frac{\hbar ^2}{2m}\frac{\pi^2}{3}n^2\left(1-\frac{4}{ \gamma}\right).\label{G-E}
\end{equation}
The low-lying excitations close to the Fermi surface have a linear
dispersion  $E=E_0+p\,v_c$ with sound velocity $v_c=v_{F}(1-{4 \gamma^{-1}})$ as $p\to 0$  
in the thermodynamic limit. Here the Fermi velocity $v_{\rm F}=2\pi n$. 
Further, from the discrete Bethe equations (\ref{BA1}) the
finite-size corrections to the ground state energy for strong coupling in the thermodynamic
limit are given by \cite{BGM}
\begin{equation}
E_0(N,L)-Le^{(\infty)}_0=-\frac{ \pi C v_c}{6L} +O(1/L^2), 
\end{equation}
with central charge $C=1$.

Now consider the 1D interacting Bose gas from the perspective of GES.
The GES parameter $\alpha_{ij}$ is defined through the linear relation 
${\Delta d_i}/{\Delta N_j}=-\alpha_{ij}$ \cite{Haldane}, i.e., 
the number of available single particle states of species $i$, 
denoted by $d_i$, depends on the number of
other species $\left\{N_j\right\}$ when one particle of species $i$ is
added.  Thus $d_i$ is given by \cite{Wu,Isakov}
\begin{equation}
d_i(\left\{N_j\right\})=G_i^0-\sum_j\alpha_{ij}N_j.
\end{equation}
Here $G_i^0 = d(\left\{0 \right\})$ is the number of available single
particle states with no particles present in the system.
The GES parameter can be determined by the two-body scattering matrix
\begin{equation}
\alpha _{ij} := \alpha (k,k')
=\delta(k,k')-\frac{1}{2\pi}\theta'(k-k'), 
\end{equation}
through counting the available states in quasimomentum space.
Here the function
\begin{equation}
\theta'(x)=\frac{2c}{c^2+x^2}.
\end{equation}
All accessible states $D_i=d_i+N_i-1$ in a momentum interval $\Delta k_i$ 
can be determined by the Bethe ansatz result
\begin{equation}
\rho+\rho_{\rm h}=\frac{1}{2\pi}+\frac{1}{2\pi}\int
_{-\infty}^{\infty}dk'\theta'(k-k')\rho(k'). \label{d-BA}
\end{equation} 
Here $\rho$ and $\rho_{\rm h}$ are the density of occupied states and
the density of holes in interval $\Delta k_i$, respectively.

The equivalence between 1D interacting bosons and ideal particles
obeying GES is clearly seen from the equivalence between the TBA \cite{Yang-Yang} result
\begin{eqnarray}
\epsilon (k)&=&\epsilon^0(k) -\mu \nonumber\\
&&-\frac{T}{2\pi}\int_{\infty}^{\infty} \!\! dk'\theta'(k-k')
\ln(1+{\mathrm e}^{-{\epsilon(k')}/{K_B T}}), \label{TBA}
\end{eqnarray}
and the GES equation \cite{Wu}
\begin{equation}
(1+w_i)\prod _j\left(\frac{w_j}{1+w_j}\right)^{\alpha  _{ji}}={\mathrm e}^{(\epsilon_i-\mu_i)/K_B T},
\label{mutual}
\end{equation}
on setting $w_i={\mathrm e}^{\epsilon(k_i)/K_BT}$. 
Here $\epsilon(k)$ is the dressed energy measuring the energy above the Fermi surface, 
with $\epsilon^0(k)=k^2$.

In the above equations, the most probable distribution is given by $\sum
_j(w_j\delta_{ij}+\beta_{ij})n_j=1$ with $n_i = N_i/G_i$ and $\beta
_{ij} = \alpha_{ij} G_j/G_i$ \cite{Wu}. 
For $\alpha=0$ the GES result (\ref{mutual}) reduces to Bose statistics.
For $\alpha_{ij}=1$ it reduces to Fermi statistics.  
The above equivalence is valid for the case $c>0$. 
We note that this type of connection between TBA and GES is very general. 
For 1D interacting systems with internal degrees of freedom the equivalence appears to be with ideal
particles obeying multi-component exclusion statistics \cite{Wu2,Fukui,Sutherland,Iguchi2}.

\section{Distribution profiles and thermodynamic properties}
\label{distribution}

In this section we concentrate on the distribution profiles and the 
thermodynamics of the strongly interacting 1D Bose gas. 
In general, the 1D interacting many-body systems have mutual GES with quasiparticle
excitations in momentum space. 
The quasiparticle excitations in these systems can be viewed as anyons with inperfect exclusion.  
In the GES description above, the state counting only involves real qusimomentum roots. 
For complex quasimomentum roots, Wu's statistics (\ref{mutual}) is not valid. 
For the 1D Bose gas, we restrict our attention to the repulsive regime, i.e., $g_{\rm 1D}>0$.  
From (\ref{mutual}) we see that at zero temperature there exists a Fermi-like surface with
energy $\epsilon_{\rm F}=\mu_0$, where $n(k)= 1/\alpha $ if
$\epsilon(k)\leq \mu_0$ and $n(k)=0$ if $\epsilon(k) > \mu_0$.
Therefore at low temperatures, a relatively small number of particles
are excited above the Fermi surface leaving an unequal amount of holes
below the Fermi surface.  
For strong coupling and at low temperatures, the majority of particles are below the surface. 
It is reasonable to expect that the statistics of the particles at finite
temperatures are equal to the statistics at zero temperature \cite{Wu2,BGO,BG}.  
Varying the repulsive interaction from weak coupling $\gamma =0^+$ to the 
strong coupling limit $\gamma \to \infty$ is equivalent to 
varying from Bose statistics $\alpha \approx 0$ to Fermi statistics with $\alpha \to 1$ 
for ideal particles obeying GES.
We see from the TBA result (\ref{TBA}) that for $c=0^+$
\begin{equation}
\epsilon(k)\approx  k^2 -\mu-\frac{T}{2\pi}\ln(1+{\mathrm e}^{-{\epsilon(k)}/{K_B T}}).
\end{equation}
Further, from the Bethe ansatz result (\ref{d-BA}) we have $\rho_{\rm h}\approx 1/2\pi$. 
Thus the distribution function is obtained as
\begin{equation}
n(k)=2\pi \rho(k)=\frac{1}{\mathrm{e}^{({k^2-\mu})/{K_B T}}-1}, 
\end{equation}
which represents the Bose statistics with $\alpha = 0$.  
This relation is easily seen from the GES result (\ref{mutual}) with $\alpha = 0$.

In the strong coupling regime and at low temperatures, the relation between the densities
$\rho$ and $\rho_{\rm h}$ is  $2\pi\left(\alpha \rho +\rho_{\rm h}\right)\approx 1$ 
with nonmutal GES determined by 
\begin{equation}
\alpha \approx 1- \frac{2}{\gamma}.\label{eq:alga}
\end{equation}
The density relation gives the most probable distribution $n(\epsilon)=2\pi \rho$ to be \cite{BG}
\begin{eqnarray}
& &n(\epsilon)=\frac{w(\epsilon)}{1+(\alpha-1) w(\epsilon)},
\label{FS-new}\\
& &\alpha\ln \left(1-w(\epsilon)\right)-\ln
w(\epsilon)=\frac{\epsilon-\mu}{K_BT}. \label{FS-2-new}
\end{eqnarray}
The relations (\ref{FS-new}) and (\ref{FS-2-new}) provide an alternative path to the 
thermodynamics of the 1D interacting Bose gas via well established results in quantum statistics. 
In particular, following Isakov {\em et al.} \cite{Isakov2} and Iguchi \cite{Iguchi}, 
at low temperatures, i.e., for $T < T_d$, where $T_d=
\frac{\hbar^2}{2m}n^2$ is the quantum degeneracy temperature, the
thermodynamics of ideal particles obeying GES can be derived from the
fractional statistics distribution $n(\epsilon)$ via Sommerfeld expansion.  
Thermodynamic properties such as the total energy $E$ and the pressure may
be calculated directly from (\ref{int}) in terms of the integral $I[f]$ given in Ref. \cite{Isakov2} 
for ideal particles obeying GES.

In order to calculate the thermodynamic properties, we write the particle number and total energy as
\begin{eqnarray}
N&=&\int_0^{\infty}d\epsilon\, G(\epsilon) n(\epsilon),\\
E&=&\int_0^{\infty}d\epsilon \,G(\epsilon) n(\epsilon)\epsilon,\label{int}
\end{eqnarray}
in terms of the density of states $G(\epsilon)$ given by
\begin{equation}
 G(\epsilon)=L/\left(2\pi
\sqrt{\frac{\hbar^2}{2m}\epsilon }\right).
\end{equation}  
In this way we find the chemical potential \cite{BG}
\begin{equation}
\mu
=\mu_0\left[1+c_2t^2+c_3t^3+c_4t^4+O(t^5)\right], \label{GES-mu}
\end{equation}
in terms of the effective temperature $t={K_BT}/{\mu_0}$,
where $\mu_0=\frac{\hbar^2}{2m}\pi^2n^2\alpha ^2$. 
The first few coefficients are given by
\begin{eqnarray}
c_2 &=& \frac{\pi^2\alpha}{12},\nonumber\\
c_3 &=&-\frac{3}{4}\zeta(3)\alpha(1-\alpha),\nonumber\\
c_4 &=& \frac{\pi^4}{144}\alpha\left(3-2\alpha+3\alpha^2\right), 
\end{eqnarray}
with the constant $\zeta(3)=\sum_{n=1}^{\infty}1/n^3 \simeq 1.20206$.

\begin{center}
\begin{figure}
\includegraphics[width=1.0\linewidth]{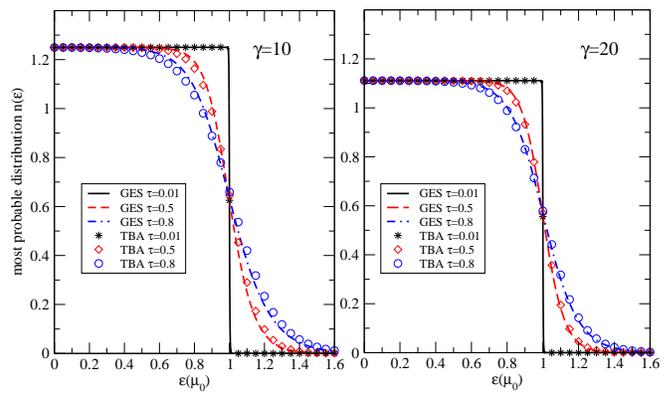}
\caption{Comparison between the GES and TBA results for the most probable 
  distribution profiles $n(\epsilon)$ for the values $\gamma =10$ (left panel) and
  $\gamma=20$ (right panel) at different values of the degeneracy
  temperature $\tau=K_BT/T_d$.  At zero temperature
  $n(\epsilon)={1}/{\alpha}$ which leads to a Fermi surface at $\epsilon=\mu_0$.
  This surface gradually decreases with increasing temperature due to 
  a large number of holes appearing below the surface. 
  Pure Fermi statistics are obtained in the limit $\gamma \to \infty$.  
  The solid lines show the most probable GES distribution for ideal particles 
  derived from (\ref{FS-new}) with (\ref{GES-mu}). 
  The symbols indicate the corresponding distributions
  evaluated from the TBA result (\ref{TBA-n}) for interacting bosons.}
\label{fig:n}
\end{figure}
\end{center}

On the other hand, for the strongly interacting Bose gas, the distribution function 
obtained from the TBA  results (\ref{d-BA}) and (\ref{TBA}) is 
\begin{equation}
n(\epsilon)=2\pi \rho(\epsilon)= 
\frac{1}{\alpha(1+{\mathrm e}^{\frac{(\epsilon-\mu)}{K_BT}})}, \label{TBA-n}
\end{equation}
with the dispersion relation $\epsilon=\frac{\hbar^2}{2m}k^2$. 
Here the chemical potential is given by
\begin{equation}
\mu =\mu_0\left[1+\frac{\pi^2}{12}t^2+\frac{\pi^4}{36}t^4 +O(t^6)\right].\label{TBA-mu}
\end{equation}
It follows that in the strong coupling limit the chemical potential (\ref{GES-mu})
derived from GES coincides with the TBA result (\ref{TBA-mu}).
For $\gamma \gg 1$ they coincide to the order shown, 
with the coefficient $c_3$ in ({GES-mu}) vanishing.
For comparison, the distribution profiles obtained from GES (\ref{FS-new}) with
(\ref{FS-2-new}) and the TBA result (\ref{TBA-n}) with (\ref{TBA-mu}) are 
shown in Figure \ref{fig:n}. 
There is good agreement between the results obtained from the two approaches 
for strongly interacting bosons.
At low temperatures the distributions for strongly interacting bosons
clearly deviate from pure Fermi statistics. 
As seen in Figure \ref{fig:n}, 
the height of the distribution function $n(\epsilon)$ approaches the
Fermi statistics value $n(\epsilon)=1$ with
increasing interaction strength $\gamma$ at zero temperature. 
During the fermionization process, i.e. as $\gamma$ increases from large but
finite values to infinity,
more than one particle is allowed to occupy a single quantum state. 
This signature is attributable to the collective behaviour of interacting bosons.

\begin{center}
\begin{figure}
\includegraphics[width=1.0\linewidth]{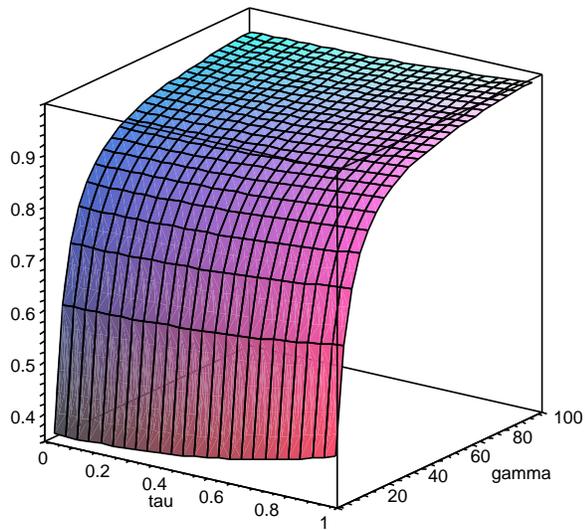} 
\caption{The total  energy per particle in units of the free Fermi energy
  $E_{\rm F}$ as a function of the effective temperature $\tau$ and the
  coupling strength $\gamma$.}
\label{fig:E}
\end{figure}
\end{center}

At low temperatures, the thermodynamic properties derived from GES
coincide with the ones derived from the TBA \cite{BG}. 
The total energy obtained from the TBA result (\ref{TBA-n}) is
\begin{eqnarray}
\frac{E}{E_0}&\approx& \left[1+\frac{1}{4\pi^2}\left(1+\frac{8}{\gamma}\right)\tau^2 \right.\nonumber\\
&&  \left. +\frac{1}{20\pi^4} \left(1+\frac{16}{\gamma}\right)\tau^4 \right],
\label{TBA-E}
\end{eqnarray}
in terms of the effective temperature $\tau=K_BT/T_d$ with degeneracy temperature  
$T_d=\frac{\hbar^2}{2m}n^2$ and
where $E_0=\frac{1}{3}N\mu_0$ is the ground state energy (\ref{G-E}) at zero temperature. 
Figure \ref{fig:E} shows a plot of the total energy per particle in units of the Fermi energy
\begin{equation} 
E_{\rm F}=\frac{\hbar^2}{2m}\frac{1}{3}\pi^2n^2, 
\end{equation}
as a function of the coupling constant $\gamma$ and the effective temperature $\tau$. 
It is apparent that the energy increases slowly as the effective temperature $\tau$ increases, with
the total energy mainly dominated by the dynamical interaction. 
The thermal fluctuations are largely suppressed. 
The energy curve approaches that obtained with Fermi statistics as 
$\gamma \to \infty$, namely in the TG limit. 

\begin{center}
\begin{figure}
\includegraphics[width=1.0\linewidth]{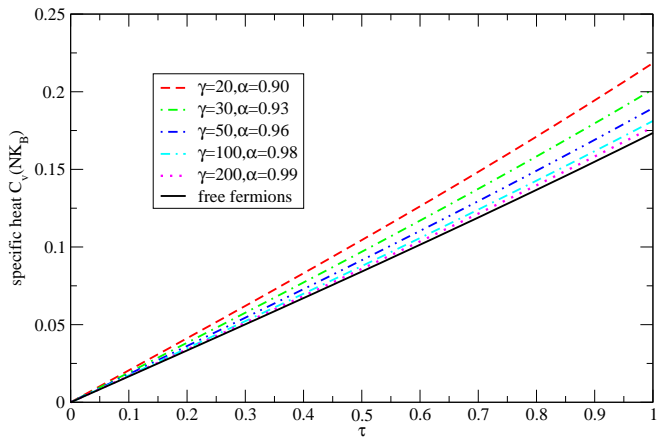}
\caption{Specific heat (\ref{TBA-Cv}) in units of $NK_B$ as a function of the degeneracy
  temperature $\tau$ for the interaction parameter values $\gamma=20,30,50,100,200,\infty$. 
  The values of the statistical parameter $\alpha$  vary with respect to $\gamma$ according to 
  (\ref{eq:alga}).}
\label{fig:Cv}
\end{figure}
\end{center}

The specific heat is given by $c_v=\left(\frac{\partial E}{\partial T}\right)_{N,L}$ with result
\begin{equation}
c_v\approx \frac16{NK_B\tau}
\left[\left(1+\frac{4}{\gamma}\right)+\frac{4}{10\pi^2}\left(1+\frac{12}{\gamma}\right)\tau^2\right].
\label{TBA-Cv}
\end{equation}
Figure \ref{fig:Cv} shows a plot of the specific heat as a function of the effective
temperature for different interaction strength.
For small temperatures the specific heat is almost linearly increasing with $\tau$.
Similar behaviour was found in the $q$-deformed Bose gas \cite{BS}. 
For strongly interacting bosons the specific heat deviates from the free Fermi curve 
as $\gamma$ decreases.  
This is mainly because the dynamical interaction $\gamma$ lowers the entropy in fermionization.  
In general the specific heat reveals an important signature of the quantum statistics 
of interacting many-body systems as it sensitively depends on the GES parameter $\alpha$.

We now consider the local pair correlations for the strongly
interacting 1D Bose gas and their role in revealing GES. 
The one-body and two-body correlation functions are key quantities 
for describing the quantum degenerate signature of 1D interacting
systems \cite{Shlyapnikov,Forrester2,CAZA2,Olshanii2,Astrak}. 
The momentum distribution and the static structure factor
can be obtained from the Fourier transform of the one-body and
two-body correlation functions, respectively. 
The local three-body correlation gives the three-body recombination rate. 
Universality in correlation functions is of particular interest in theory
\cite{CAZA1,Shlyapnikov,Forrester2,CAZA2,Olshanii2,Astrak,Haldane2} 
and experiment \cite{E-TG2,Weiss1}. 
In particular, the local pair correlations for the 1D Bose gas
have been observed experimentally by measuring photoassociation rates \cite{Weiss1}.
Physically, the local pair correlation is a measure of the probability of observing 
two particles in the same place.

\begin{center}
\begin{figure}
\includegraphics[width=1.0\linewidth]{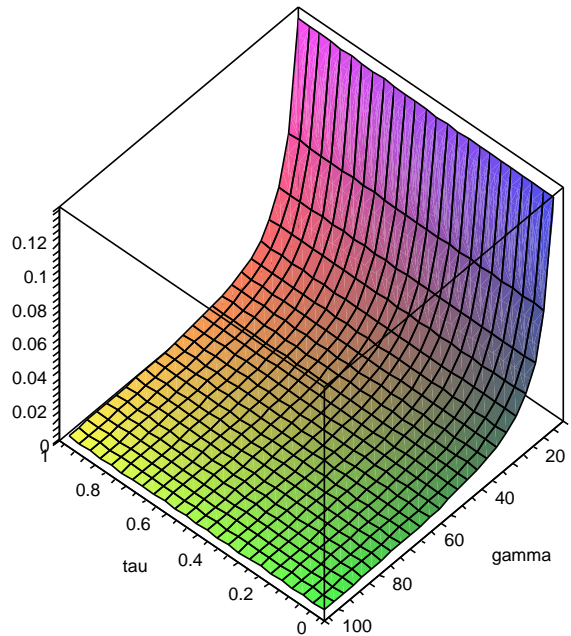} 
\caption{The local pair correlation $g^{(2)}(0)$ as a function of the effective
  temperature $\tau$ and the coupling strength $\gamma$.}
\label{fig:g2}
\end{figure}
\end{center}

In the grand canonical description, the two-particle local correlation is given 
by \cite{KOR,Shlyapnikov,CAZA2}
\begin{equation}
g^{(2)}(0)=\frac{2m}{\hbar^2n^2}\left(\frac{\partial f(\gamma,\tau)}{\partial \gamma }\right)_{n,\tau},
\end{equation}
where $f=\mu-2E/N$ is the free energy per particle. 
In this way the result   
\begin{equation}
g^{(2)}(0)\approx
\frac{4\pi^2}{3\gamma^2}\left(1+\frac{\tau^2}{4\pi^2}+\frac{3\tau^4}{80\pi^4}\right),
\label{eq:g2}
\end{equation}
for the two-particle local correlation at low temperatures follows from the TBA.
Figure~\ref{fig:g2} shows the local pairing correlation as a function of the
effective temperature $\tau$ and the interaction strength $\gamma$. 
We see that the dynamical interaction dramatically reduces the pair correlation
due to decoherence between individual wave functions. 
It tends to zero as $\gamma \to \infty$, namely in the TG limit.
Moreover, the pair correlation is only weakly dependent on the 
temperature because the thermal fluctuations are suppressed at the
temperatures $T\ll T_{\rm d}$.
The experimental values for the local pair correlations \cite{Weiss1} 
are in excellent agreement with the theoretical result (\ref{eq:g2}) for 
over a wide range of interaction strength for $\tau \approx 0$.

\section{Discussion}
\label{conclusion}

We have demonstrated that the statistical profiles of the strongly
interacting 1D Bose gas at low temperatures are equivalent to those of a gas of 
ideal particles obeying nonmutual GES.
For the strong coupling limit $\gamma \to \infty$, the TG gas, the equivalence
is with Fermi statistics.
The deviation from Fermi statistics during the fermionization process is described by
the relation $\alpha \approx 1 - 2/\gamma$, where $\alpha$ is a measure of the GES.
In the strong coupling regime the local pair correlation function (\ref{eq:g2}) may also be written as
\begin{equation}
g^{(2)}(0)\approx
\frac{\pi^2}{3}(1-\alpha)^2 \left(1+\frac{\tau^2}{4\pi^2}
+\frac{3\tau^4}{80\pi^4}\right).
\label{eq:g2alpha}
\end{equation}
This suggests the possibility of observing the quantum degenerate
behaviour of ideal particles obeying GES through experiments on 1D
interacting systems.
In this sense the recent experimental measurement of the local pair correlation function
 in a 1D Bose gas of $^{87}$Rb atoms \cite{Weiss1} also provides a measurement of GES.
In particular, the deviation of the GES parameter $\alpha$ away from the 
pure Fermi statistics value $\alpha=1$.
This gives a quantitative description of the fermionization process.
Other thermodynamic properties, such as the distribution profiles depicted
in Figure 1 and the specific heat, are also sensitive to the statistics.
They also thus provide a way of exploring fractional statistics in the 
strongly interacting 1D Bose gas.

\noindent
{\em Acknowledgments.}
This work has been supported by the Australian Research Council.  
The authors thank M. Bortz and C. Lee for helpful discussions on the 1D Bose gas.
MTB also thanks the organizers of the Scientific Seminar on Physics of Cold Trapped Atoms 
at the 15th Laser Physics Workshop for the opportunity to present our results.

\end{document}